\documentclass[12pt]{article}

\def\vkp#1{\vec{k}_{#1\perp}}
\def\vq#1{\vec{q}_{#1}}
\def\cO#1{{\cal{O}}\left(#1\right)}

\def\MSbar{\overline{\mbox{MS}}}

\newskip\humongous \humongous=0pt plus 1000pt minus 1000pt
\def\caja{\mathsurround=0pt} \def\eqalign#1{\,\vcenter{\openup1\jot
\caja   \ialign{\strut \hfil$\displaystyle{##}$&$
\displaystyle{{}##}$\hfil\crcr#1\crcr}}\,} \newif\ifdtup

\renewcommand{\theequation}{\thesection.\arabic{equation}}
\newcommand\mysection[1]{\section{#1} \setcounter{equation}{0}}

\newcounter{hran}
\renewcommand{\thehran}{\thesection.\arabic{hran}}

\def\bmini{\setcounter{hran}{\value{equation}}
\refstepcounter{hran}\setcounter{equation}{0}
\renewcommand{\theequation}{\thehran\alph{equation}}\begin{eqnarray}}

\def\bminiG#1{\setcounter{hran}{\value{equation}}
\refstepcounter{hran}\setcounter{equation}{-1}
\renewcommand{\theequation}{\thehran\alph{equation}}
\refstepcounter{equation}\label{#1}\begin{eqnarray}}

\def\emini{\end{eqnarray}\relax\setcounter{equation}{\value{hran}}\renewcommand{\theequation}{\thesection.\arabic{equation}}}

\def\kps{k_\perp^2}
\def\frac#1#2{ {{#1} \over {#2} }}

\def\half{\mbox{\small $\frac{1}{2}$}}

\def\VEV#1{\left\langle #1\right\rangle}

\def\ie{\hbox{\rm i.e. }}

\def\fm{\,{\rm fm}}
\def\ae{\alpha_{\mbox{\scriptsize eff}}}
\def\cR{{\cal{R}}}
\def\cS{{\cal{S}}}
\def\cH{{\cal{H}}}
\def\cM{{\cal{M}}}
\def\e{\mathrm{e}}

\def\fun#1#2{\lower3.6pt\vbox{\baselineskip0pt\lineskip.9pt
  \ialign{$\mathsurround=0pt#1\hfil##\hfil$\crcr#2\crcr\sim\crcr}}}

\def\abs#1{\left| \: #1 \: \right|}%

\def\beq{\begin{equation}}
\def\eeq{\end{equation}}
\def\beql#1{\begin{equation}\label{#1}}
\def\beeq{\begin{eqnarray}}
\def\eeeq{\end{eqnarray}}
\def\bit{\begin{itemize}}
\def\eit{\end{itemize}}

\def\bp{\bar p}

\def\as{\alpha_{\mbox{\scriptsize s}}}

\def\np#1#2#3{Nucl.\ Phys.\ B#1 (19#3) #2}
\def\pl#1#2#3{Phys.\ Lett.\ #1B (19#3) #2}
\def\pr#1#2#3{Phys.\ Rev.\ D #1 (19#3) #2}

\def\prep#1#2#3{Phys.\ Rep.\ #1 (19#3) #2}
\def\prl#1#2#3{Phys.\ Rev.\ Lett.\ #1 (19#3) #2}

\def\spj#1#2#3{{\em Sov.Phys.JETP}\/~\underline{#1} (19#3) #2}

\begin{document}
\begin{titlepage}
\renewcommand{\thefootnote}{\fnsymbol{footnote}}
\begin{flushright}
     IFUM-573-FT \\
     hep-ph/9707532 \\
     July 1997 \\
\end{flushright}
\par \vskip 10mm
\begin{center}
{\Large \bf
Universality of $1/Q$ corrections to jet-shape observables rescued\footnote{
Research supported in part by MURST, Italy
and by EC Programme ``Human Capital and Mobility",
contract CHRX-CT93-0357 (DG 12 COMA).}}
\end{center}
\par \vskip 2mm
\begin{center}
{\bf
Yu.L. \ Dokshitzer\footnote{
On leave from St. Petersburg Nuclear Institute,
Gatchina, St. Petersburg 188350, Russia},
A.\ Lucenti,  G.\ Marchesini and G.P.\ Salam}\\
\vskip 5 mm
Dipartimento di Fisica, Universit\`a di Milano \\
and INFN, Sezione di Milano, Italy
\end{center}

\par \vskip 2mm
\begin{center} {\large \bf Abstract} \end{center}
\begin{quote}
  We address the problem of potential non-universality of the leading
  $1/Q$ power corrections to jet shapes emerging from the
  non-inclusive character of these observables.  
  We consider the thrust distribution as an example and analyse the 
  non-inclusive contributions which emerge at the two-loop level.
  Although formally subleading in $\as$,
  they modify the existing na{\"\i}ve one-loop result for the expected
  magnitude of the power term by a factor of order unity.  
  Such a promotion of a subleading correction into a numerical factor is
  natural since the non-perturbative power terms are explicitly proportional 
  to powers of the QCD scale $\Lambda$ which can be fixed precisely only at 
  the two-loop level. 
  The ``jet-shape scaling factor'' depends on the observable but remains
  perturbatively calculable.  Therefore it does not undermine the 
  universal nature of $1/Q$ power corrections, which remain expressible
  in terms of the universal running coupling and universal
  soft-gluon emission. 
\end{quote}
\end{titlepage}

\mysection{Introduction}
The analysis of power-behaved contributions to infrared and collinear safe
characteristics of hard QCD processes has recently been
developed as a method to quantify non-perturbative 
effects~[\ref{Web94}--\ref{NS}].
It has been recognised that the leading power corrections to 
infrared/collinear safe quantities 
are determined by an incomplete cancellation of soft-gluon contributions.
A QCD process at a hard scale $Q\gg\Lambda$ is determined by the small
space-time intervals between the quarks and gluons involved.  In spite
of that, at the level of power suppressed corrections, the cross section
of such a process may acquire a contribution from a soft gluon which
travels a finite distance $1\fm\gg 1/Q$
and thus is sensitive to the non-perturbative interaction domain.

The sensitivity of a given observable to soft-gluon radiation 
can be studied perturbatively to determine the
{\em power}\/ $n$ of the expected leading $Q^{-n}$ correction.
Such theoretical input is valuable on its own, especially for jet physics
which deals with intrinsically Minkowskian quantities and therefore 
typically receives no hints from the Euclidean-based OPE technology 
[\ref{OPE}].  
For example, an analysis of large-distance soft-gluon effects predicts 
that the properties of jets assembled with the use of the (once
standard) JADE jet finder are contaminated by confinement effects at
the level of 
$Q^{-1}$ corrections, while the Durham ($k_\perp$) algorithm produces jets
whose rates will depart from perturbative predictions only at the 
$Q^{-2}$ level~[\ref{Web94}].

More challenging is the prediction of the {\em magnitude}\/ of expected 
power corrections. 
For a single observable this would be impossible given the present 
state of the art. 
Nevertheless it is natural to expect that soft-gluon universality 
implies proportionality between the magnitudes of power-behaved terms 
in different observables~[\ref{StKoZakh}].
Along these lines one aims to predict the relative sizes 
of power corrections to observables belonging to the same class, 
linking together, for example, jet mass and thrust
distributions~[\ref{DW97}] 
with $1/M_Q$ effects in heavy quark energy spectra~[\ref{DKT},\ref{NW}] 
($n\!=\!1$)
or power corrections to the Drell-Yan
$K$-factor~[\ref{StKoZakh},\ref{BB95}] 
with those in the 
Gross-Llewellyn-Smith DIS sum rule and $e^+e^-$ fragmentation 
functions~[\ref{BPY}] ($n\!=\!2$).

However, some doubt has been expressed as to the universality of power
corrections to jet shapes because the latter are not fully
inclusive with respect to final-state branching of the soft gluon
under focus.  To the best of our knowledge this question was
first considered by Nason and Seymour in~[\ref{NS}].

The notion of soft-gluon universality contains two ingredients.
First, it exploits the universal character of the soft-radiation
matrix element in the spirit of Low's theorem~[\ref{LBK}]. 
Secondly, it requires universality of the gluon radiation intensity 
(running coupling).

The running QCD coupling in hard processes emerges after inclusive
integration over the gluon decay products,
resulting typically in $\as(k^2_\perp)$ with $k_\perp$ the gluon
transverse momentum~[\ref{kt}].
In QED it suffices to consider the virtual photon self-energy 
blobs (``bubbles''), while in the QCD context there are also the
corrections to the gluon emission vertex.
The dispersive representation has been used in~[\ref{BPY}] 
to give meaning to the QCD running coupling 
at low scales~[\ref{DKT},\ref{lowscales}].

The value of a jet-shape variable depends, however, on the kinematics
of all final state particles (hadrons, partons). 
As long as the observable is collinear safe the (quasi-)collinear gluon decays,
which are the main contributors to the running $\as$, do not affect
the observable and therefore can be treated inclusively.  
However, in the ``large-angle'' gluon decay region, the inclusive 
interplay between real and virtual gluon loops gets broken because 
a real decay may correspond to a value of the observable different 
from that for the case of a virtual loop.
These contributions seem to undermine the above-mentioned universality.

In the case of the thrust distribution, for example, the na{\"\i}ve inclusive
treatment applies as long as the decay products fall into the same
hemisphere.  Then the contribution to thrust, $T$, can be attributed to the
parent gluon, and integration over the virtuality of the latter
results in the running coupling in a standard way.

The existence of the part of the decay phase space 
in which the two gluon offspring partons (quarks or gluons) 
happen to fly into opposite hemispheres, 
has two consequences.
First, this region has to be treated separately as its contribution
to $T$ depends on details of the gluon decay matrix element.
At the same time the effective intensity of emission of the parent gluon 
as a whole gets modified.
This is due to the fact that to reconstruct the running coupling one
sums inclusively over the gluon decay products, while in this case one
modifies the final state phase space. Thus the running coupling 
(its ``spectral density'') loses part of its standard support.

It has been argued that such a non-inclusive contribution is higher order in
$\as$. 
Its quark-antiquark (``Abelian'') part was studied numerically 
in [\ref{NS}] and found to be small.

In this paper we analyse the question of (non-)universality of the
leading power corrections to jet shapes and thrust in particular. 
We find that the na{\"\i}ve inclusive treatment has to be modified 
by two effects.
The first is due to non-inclusive configurations of the parent gluon decay
(non-inclusive correction). The second is due to an incomplete compensation 
between real and virtual logarithmic terms coming from the soft-gluon 
splitting into gluons (inclusive correction). This correction is
absent in the contribution from the gluon splitting into a
quark-antiquark pair. 
We show that both non-inclusive and inclusive corrections, 
although formally subleading in $\as$, get promoted to a factor 
{\em of order unity}\/  in the magnitude of the $1/Q$ power correction.
As a consequence, previously obtained results 
based on the na{\"\i}ve inclusive treatment of an observable $j$
acquire a ``rescaling factor'' $(1+r_j)$
which is {\em calculable}\/ but depends on the observable under consideration.

In this sense one can still consider power terms
in jet observables as universal, i.e. based on the universal coupling 
and the universal soft-gluon radiation pattern, although 
their relative magnitudes differ from the na{\"\i}ve expectations based on
one-loop off-shell soft-gluon matrix elements.
 
For thrust values close to 1, only soft-gluon radiation is essential. 
For $T$ well below 1, one has contributions both 
from multiple soft-gluon emission and from rare $N$-jet configurations 
($N>2$). 
The first contribution has been found~[\ref{DW97}] to be dominant
at least down to $ T=2/3$, 
the minimum $T$ for any three-jet configuration. 
As far as power corrections are concerned, 
it was shown in~[\ref{NS}] that the order $\as(Q^2)$ three-jet configuration 
does not develop a $1/Q$ power correction. 

Therefore to analyse the thrust distribution at two loops we take into
consideration [\ref{DMO}] particle ensembles consisting of the primary 
quark-antiquark pair
accompanied by partons originating from multiple emission of primary
soft gluons and employ the existing all-order resummed 
perturbative prediction~[\ref{CWT}].

We extract the leading $1/Q$ power term in the
radiator, namely the exponent of the Mellin-transformed 
$T$-distribution. 
We do this in the framework of the treatment of 
non-perturbative power corrections to infrared/collinear safe observables 
introduced in~[\ref{BPY}], based on the
notion of an infrared-finite strong coupling which 
differs from the perturbative form in the infrared region.  

At the two-loop level, non-inclusiveness of the thrust reveals itself
and the characteristic QCD scale $\Lambda$ becomes precisely defined 
through the fixing of the scheme for the running coupling. 
To this end we choose the CMW renormalisation scheme~[\ref{CMW}], 
in which $\as$ is defined as the intensity of soft-gluon radiation.

Section 2 contains the necessary ingredients for the two-loop analysis
of the radiator of the thrust distribution: kinematics, 
structure of the soft-gluon emission followed by
gluon splitting into $q\bar{q}$ and gluon pairs,  
scheme fixing, and thrust resummation.
In section 3 we recall the dispersive method of [\ref{BPY}]
and its application to the thrust distribution in the na{\"\i}ve 
inclusive approach [\ref{DW97}].
Section 4 is devoted to the exact treatment of the problem,
the analysis and calculation of the inclusive and 
non-inclusive contributions to the thrust rescaling factor $(1+r_T)$.  
Section 5 contains a discussion of the results and the prospects 
for the future.

\mysection{Two-loop analysis of thrust distribution}
\subsection{Thrust: kinematics}
Consider a QCD hard process in which the final state consists of
a leading quark-antiquark pair with momenta $p$ and $\bar p$
together with a system of $N$ partons (quarks and gluons) 
with momenta $k_1, \cdots , k_N$.
It is convenient to cast the cross section in terms of the Sudakov
(light-cone) variables defined with respect to the thrust axis of the event.
To this end we introduce two light-like vectors
$$
 P^\mu\>, \>\> \bar{P}^\mu\>; \quad P^2=\bar{P}^2=0\>; \quad
 P^\mu + \bar{P}^\mu= Q^\mu\>,\>\> 
 2(P\bar{P}) = Q^2\>\equiv\>1\>, 
$$
and represent the parton 4-momenta as
$$
 k_i^\mu \>=\> \beta_i P^\mu \>+\> \alpha_i \bar{P}^\mu \>+\>
 (k_{i\perp})^\mu,
 \qquad\qquad \alpha_i\beta_i = k_{i \perp}^2
$$
The leading quark momenta become
$$\eqalign{
p&= (1-\sum_{i=1}\beta_i-\beta_{\bar{p}})P\>+\>\alpha_p\bar{P}\>+\>
p_\perp\>, \cr 
\bar{p}&=(1-\sum_{i=1}\alpha_i-\alpha_p)\bar{P}
 \>+\> \beta_{\bar{p}} P\>+\> \bar{p}_\perp\>,
}$$
Adding the normalised longitudinal momenta of the secondary partons,
$
\left. k_{i||} \right/Q = \abs{\alpha_i-\beta_i}/2\>,
$ 
with those of the two quarks,
for the thrust value we obtain
\beql{Ti}\eqalign{
 T &= \sum_i\half\abs{\alpha_i-\beta_i} 
+ \half[\, (1-\sum_{i=1}\beta_i-\beta_{\bar{p}})-\alpha_p\,]
+ \half[\, (1-\sum_{i=1}\alpha_i-\alpha_p)-\beta_{\bar{p}}\,] \cr
&= 1-\sum_i \min\{\alpha_i,\beta_i\} - (\alpha_p+\beta_{\bar{p}})\>.
}\eeq
This expression implies that $p$ and $\bar{p}$ momenta belong to opposite
hemispheres, which is the dominant configuration 
(the correction is numerically small and relatively suppressed 
{\em at least}\/ as $(1\!-\!T)$ when $T$ is large). 

The values of the ``small components'' of the quark momenta, 
$\alpha_p$ and $\beta_{\bar{p}}$ depend on quark transverse momenta
with respect 
to the thrust axis through the on-mass-shell conditions,
$p^2=\bar{p}^2=0$. 
Transverse quark recoils are typically of order
$$
   p_\perp^2\sim \bar{p}_\perp^2 \sim \sqrt{N}\cdot k_{\perp}^2\>,
$$
with $k_\perp$ characteristic transverse momentum of a secondary parton.
We have kept in the estimate a (logarithmic) enhancement factor due 
to the parton multiplicity $N$.

As we shall see the leading power correction comes from the region
$\alpha_i \sim \beta_i$ of the order of the transverse momentum 
(in units of $Q$). 
Therefore in (\ref{Ti}) we shall neglect $\alpha_p$ and $\beta_{\bar p}$ 
which are of the order of the transverse momentum squared. 

\subsection{Gluon splitting}
\paragraph{Two-body variables and phase space.}
The two-parton phase space is
$$
  d\Gamma_2(k_1,k_2) = (4\pi)^4 \frac{d^4k_1\,d^4k_2}{(2\pi)^6} \,
\delta(k_1^2)\delta(k_2^2)
= \frac{4}{\pi^2}  
\left(\frac12 \frac{d\alpha_1}{\alpha_1}\,{d^2k_{1\perp}} \right) 
\left(\frac12 \frac{d\alpha_2}{\alpha_2}\,{d^2k_{2\perp}} \right) 
\>.
$$
Introducing the relative fraction of $\alpha$-components, $z$, and 
the invariant mass squared of the two-parton system $m^2$
we write:
$$
\eqalign{
  \alpha &\equiv\alpha_1+\alpha_2\>, \quad 
   \alpha_1=z\alpha\>, \>\> \alpha_2=(1-z)\alpha\>, \cr
  m^2 &= (k_1+k_2)^2 = (\alpha_1\beta_2+\alpha_2\beta_1)-
  2\vec{k}_{1\perp} \cdot \vec{k}_{2\perp}
 = z(1-z) 
\left[\,\frac{\vec{k}_{1\perp}}{z} - \frac{\vec{k}_{2\perp}}{1-z}\,\right]^2.
}$$
In terms of the scaled parton transverse momenta,
$$
\vq{1}=\frac{\vkp{1}}{z}\>, \quad \vq{2}=\frac{\vkp{2}}{1\!-\!z}\>,
\qquad \vq{} = \vq{1} - \vq{2},
$$ 
the phase space takes the form
\beql{d2Gamma}
  d\Gamma_2(k_1,k_2) = 
 \frac{d\alpha}{\alpha}\, {dz}\, \frac{d^2q_{1}}{\pi}
  \frac{d^2q_{2}}{\pi}\>  dm^2\,\delta\left(\frac{m^2}{z(1\!-\!z)}-
(\vq1-\vq2)^2 \right).
\eeq
For later use we also define
\beq\eqalign{
 \vec{k}_\perp &= \vkp1 + \vkp2 \>=\> z\vq1 + (1-z)\vq2 \>; \cr
 \beta & = \beta_1+\beta_2 = \frac{k_\perp^2+m^2}{\alpha}\>, \quad 
   \beta_1=\frac{z}{\alpha}q_1^2\>; \>\>
   \beta_2=\frac{1\!-\!z}{\alpha}q_2^2 \>.
}\eeq

\paragraph{Matrix elements.}
The squared matrix elements for the production of two gluons or a
quark pair 
with 4-momenta $k_1$, $k_2$ read, in the soft limit $(Qk_i)\ll Q^2$
(see [\ref{DMO}])
\beql{Mrr2} \eqalign{
M^2(k_1,k_2) &=\>  4C_F^2\;W_1W_2 \>+\>4C_F\left( M^2_{gg}+M^2_{qq}\right) \cr
4\left( M^2_{gg}+M^2_{qq}\right) &= \> 
{C_A}\;(2S+H_{g}) \>+\>  
{n_f}\, H_q\>.
}\eeq
The first term, proportional to $C_F^2$, describes the independent
(Abelian) emission of two soft gluons off the hard $q\bar{q}$, where
$$
W_i \>\equiv\>   W(k_i) \>=\> \frac{p\bp}{(pk_i)(k_i\bp)} 
= \frac{2}{k^2_{i\perp}}
$$
is the standard dipole radiation cross section.

The second (non-Abelian) term, proportional to $C_F\,C_A$,
contains two pieces. 
The $S$ term describes the leading infrared contributions which is singular
in the ratio of gluon energies:
$$
S \equiv W_{12}+W_{21}-W_1\;W_2 \>, 
$$ 
where
$$
W_{ij} \equiv \frac{(p\bp)}{(pk_i)(k_ik_j)(k_j\bp)} 
\>=\> \frac{4}{m^2}\,\frac{1}{\alpha_i\beta_j}
$$
stands for the ordered 2-dipole emission.

$H_g$ and $H_q$ are responsible for parton configurations with energies of
the same order. 
The ``hard'' parts of the two-gluon ($H_g$)
and the two-quark ($H_q$)
contributions read
\bminiG{Hgq}
\label{Hg}
H_g &\equiv& -SR + J^2 - 4\frac{W(k)}{(k_1k_2)} \>, \\
\label{Hq}
H_q &\equiv& -J^2 + 2\frac{W(k)}{(k_1k_2)} \>.
\emini
Here
\beq\eqalign{
\label{J2R}
 J^2 &\>=\> 2 \left(
 \frac{(pk_1)(k_2\bp)-(pk_2)(k_1\bp)}{(k_1k_2)(pk)(k\bp)}
 \right)^2
 = 8\left(\frac{\alpha_1\beta_2-\alpha_2\beta_1}
    {m^2\,\alpha\beta}\right)^2 \>, \cr
R &\>=\> \frac{(pk_1)(k_2\bp)+(pk_2)(k_1\bp)}{(pk)(k\bp)}\>
=\> \frac{\alpha_1\beta_2+\alpha_2\beta_1}{\alpha\beta}\>,
}\eeq
and $W(k)= {(p\bp)}/{(pk)(k\bp)}$ is 
the dipole emission of $k=k_1+k_2$.

We may write the squared matrix elements in the following form:
\beql{cMreduced}
 M^2_{gg}+M^2_{qq} \>=\>
 \frac{\cM^2}{m^2(zq_1^2+(1-z)q_2^2)}\>,
\eeq
where
\beql{cMsq}
\cM^2 \>=\>  2 C_A \left(2\cS +\cH_g\right) + 2n_f \cH_q.
\eeq
The distributions determining $\cM^2$ are functions of the 
dimensionless ratios $u_i^2$ with 
$$
  \vec{u}_i = \frac{\vec{q}_i}{|q|}\>.
$$
Namely,
\bminiG{matelms}
\label{matelmgg}
2\cS &=& \frac{zu_1^2+(1\!-\!z)u_2^2}{z(1\!-\!z)}\left(\frac1{u_1^2}
 +\frac1{u_2^2}-\frac1{u_1^2u_2^2}\right) \>,\\
\cH_g &=& -2 +z(1\!-\!z)\frac{(u_1^2-u_2^2)^2}{zu_1^2+(1\!-\!z)u_2^2}
-\frac{u_1^2+u_2^2}{2}
 \left[\,\frac1{u_1^2}+\frac1{u_2^2}-\frac1{u_1^2u_2^2} \,\right] ,\\
\cH_q &=& 1-z(1\!-\!z)\frac{(u_1^2-u_2^2)^2}{zu_1^2+(1\!-\!z)u_2^2}\> .
\emini
The collinear limit ($q^2\propto m^2\to 0$) corresponds to taking
$u_1^2\approx u_2^2\to \infty$ while keeping fixed the modulus of the
vector difference,
$$
\left(\vec{u}_1-\vec{u}_2 \right)^2=1\>, 
$$
see (\ref{d2Gamma}).
Observing (see Appendix A) that the azimuthal average of the ratio
$$
 \VEV{\frac{(u_1^2-u_2^2)^2}{zu_1^2+(1\!-\!z)u_2^2}} \>\to\> 2\>,
$$
we derive the parton splitting functions
\bminiG{splitfs}
\label{splitgg}
  2C_A(2\cS+\cH_g) &\to& 4C_A\left(
    \frac{1}{z(1-z)}-2+z(1-z)\right), \\ 
\label{splitqq}
  2n_f \cH_q &\to& 2n_f\left(1-2z(1-z)\right)\>.
\emini
Note that the latter expression is twice the standard $g\to q\bar{q}$
splitting probability because thrust as an inclusive observable does not
differentiate between quarks and antiquarks, and the symmetry factor
$1/2!$ will be included in the thrust trigger-function, see (\ref{OT})
below.

\subsection{Gluon decay and the $\beta$-function}
The $\beta$-function is related to
the integrated two-parton production probability.
Keeping fixed the invariant mass and total transverse momentum 
of the pair, and integrating $M^2$ over the azimuthal angle
$\phi$ with respect to $\vkp{}$ and over the relative momentum
fraction $z$, it is straightforward to derive
(see Appendix~A)
\beql{beta0}
 \frac1{2!}\,\int_0^1 dz  \int_0^{2\pi} \frac{d\phi}{2\pi}
\left(M_{gg}^2+M_{qq}^2\right)
= \frac{1}{m^2(k_\perp^2+m^2)} \left(-\beta_0
+2 C_A \ln\frac{k_\perp^2(k_\perp^2+m^2)}{m^4}\right)\>,
\eeq
where
\beql{betadef}
 \beta_0 = \frac{11}3N_c-\frac{2}{3}n_f \>.
\eeq
The logarithmic singularity at $m^2\to0$ comes from the soft
singularity in the $g\to gg$ matrix element (splitting function)
and is cancelled by the corresponding
virtual correction to the one-gluon radiation vertex.
The latter can be written as an integral over an intermediate gluon
virtuality $\mu^2$,
\beql{chidef}
 \chi \>=\> -\int_0^\infty \frac{d\mu^2\> k_\perp^2}
 {\mu^2(k_\perp^2+\mu^2)} 
 \left\{ 2C_A \ln\frac{k_\perp^2(k_\perp^2+\mu^2)}{\mu^4}\>+\> 
 s\left(\frac{\mu^2}{k_\perp^2}\right) \right\} 
 \left(\frac{\as}{4\pi}\right)^2.
\eeq
Here we have chosen the form of the logarithmic term so as to cancel
exactly the corresponding real emission contribution in (\ref{beta0}).
The mismatch function $s(x)$ vanishes at the origin, $s(0)=0$, and
depends on the scheme one adopts for calculating the divergent vertex
correction.  

\subsection{Fixing the scheme for $\as$}    

To fix the scheme for defining the perturbative expansion parameter
$\as$, we have to consider first the relation between particle
production at order $\as^2$ and the coupling. 
To this end we address the non-Abelian contribution to an inclusive
quantity such as the quark anomalous dimension. 

The inclusive probability for producing a soft gluon or a two-parton system
with gluon quantum numbers reads
\beq
 dw=4C_F\frac{d\alpha}{\alpha}\, \frac{d\kps}{\kps}\> \gamma(\as)\,,
\eeq
where the ``anomalous dimension'' $\gamma$ becomes
\beq
 \gamma \>=\>  \frac{\alpha_0}{4\pi}
 + \chi(k_\perp^2)
 + \int_0^\infty \frac{dm^2\> \kps}{m^2(\kps+m^2)}
   \left(\frac{\as}{4\pi}\right)^2 \left\{ -\beta_0
 + 2C_A \ln\frac{\kps(\kps+m^2)}{m^4} \right\}.
\eeq
The singular logarithmic term cancels against that in $\chi(k_\perp^2)$,
and we arrive at
\beql{gammaalpha}
 \gamma \>=\>  
 \frac{\alpha_0}{4\pi}
 + c_s   \left(\frac{\as}{4\pi}\right)^2
 + \int_0^\infty \frac{dm^2\> \kps}{m^2(\kps+m^2)}
   \left(\frac{\as(m^2)}{4\pi}\right)^2  (-\beta_0)\>.   
\eeq
Here $c_s$ is a constant given by the integral of the scheme-dependent  
piece of $\cR_{12}$, see (\ref{chidef}):
$$
   c_s = \int_0^\infty \frac{dx}{x(1+x)}\, s(x)\>.
$$
First we observe that, to the necessary accuracy, $\gamma$ satisfies, 
$$
 \kps\frac{d}{d\kps}\, \gamma \>=\> 
 -\beta_0 \left(\frac{\as(\kps)}{4\pi}\right)^2 +\ldots \>\simeq\>
  \kps\frac{d}{d\kps}\,\frac{\as(\kps)}{4\pi}\>.
$$
This means that we can choose $\gamma$ to represent the running 
coupling $\as$. Equation (\ref{gammaalpha}) then becomes a
dispersive relation for $\as(\kps)$, 
in which the combination of $\alpha_0$ and $c_s\as^2$ 
defines the coupling constant at $\kps=0$:
\beql{andim}
\gamma\equiv \frac{\as(\kps)}{4\pi} \>=\> \frac{1}{4\pi}
\left\{ \as(0) \>+\> 
\int_0^\infty \frac{dm^2\> \kps}{m^2(\kps+m^2)}\>\rho_s(m^2)\right\}
\>=\> -\frac{1}{4\pi}
\int_0^\infty \frac{dm^2}{\kps+m^2}\>\rho_s(m^2),
\eeq
where $\rho_s(m^2)$ is the spectral density, to be discussed later.

This corresponds to choosing the CMW renormalisation scheme [\ref{CMW}]
in which
$\as$ is defined as the intensity of soft-gluon radiation.
In another scheme, e.g. $\MSbar$, the anomalous dimension in the soft
limit would contain both $\as$ and $\as^2$ terms.

\subsection{Thrust: resummation}

For a $q\bar{q}+N$-parton final state the thrust distribution becomes 
\begin{equation}
   \frac{d\sigma}{dT} \>=\> \int d\Gamma_N \> M^2(q_1, q_2, \ldots
   q_N)
\> \delta(1-T-\sum_{i=1}^{N} \min\{\alpha_i,\beta_i\}) 
\>,
\end{equation}
with $M$ the $N$-parton matrix element and $d\Gamma_N$ the
corresponding $N$-body phase space.
In the soft approximation ($1-T\ll 1$) the final partons are either
relatively soft gluons emitted off the quark-antiquark line (primary
gluons) or their decay products.
The matrix element for an ensemble of primary gluons is very simple in
the soft approximation, as it is given by the product of independent
radiation amplitudes (essentially an Abelian pattern).
By taking a Mellin representation for the $\delta$-function,
$$
 \delta\left(1-T-\sum_{i=1}^{N} \min\{\alpha_i,\beta_i\}\right) 
 = \int \frac{d\nu}{2\pi i} \exp\{\nu(1-T)\} \prod_{i=1}^{N} 
 \exp\{-\nu \min\{\alpha_i,\beta_i\}\}\>,
$$
the thrust distribution takes the form
\beq
   \frac1{\sigma}
   \frac{d\sigma}{dT} \>=\> \int \frac{d\nu}{2\pi i} \>\e^{\nu(1-T)}
   \exp\{ {\cal{R}}(\nu) \}\>.
\eeq
At next-to-leading order, the ``radiator'' contains two terms,
$\cR(\nu)= \cR_1+ \cR_2$, due respectively to 
one- and two-parton contributions.

\paragraph{One-gluon contribution.} 
The one-gluon contribution, $\cR_1$, is given by the following formal 
expression: 
\beq
\label{cR1}
 \cR_{1} \>=\> 8C_F\int_0^1 \frac{d\alpha}{\alpha}
 \int\frac{dk_\perp^2}{k_\perp^2}\>
\left(\e^{-\nu\alpha}-1\right)   \Theta(k_\perp^2-\alpha^2)
\cdot \left( \frac{\alpha_0}{4\pi} \>+\> \chi(\kps) \right) ,
\eeq
where the $\Theta$-function selects the smaller longitudinal component
($\alpha<\beta=k_\perp^2/\alpha$).
The term with $\exp(-\nu\alpha)$ corresponds to the case in which the 
gluon is emitted, while $-1$ corresponds to the case in which the
gluon is in a virtual loop.
The function $\chi(k_\perp^2)$ is the order $\as^2$ contribution that
describes the $k_\perp$-dependent one-loop vertex correction to 
one-gluon emission, given in (\ref{chidef}).

Using the expression for the virtual correction, (\ref{chidef}),
we can split $\cR_1$ into two contributions
$$
   \cR_1\>=\>  \cR_{11}\>+\> \cR_{12} \>,
$$
where
\bminiG{cR1112}
\label{cR11}
 \cR_{11} \!\!&=&\!\! 8C_F\int_0^1 \frac{d\alpha}{\alpha}
 \int\frac{dk_\perp^2}{k_\perp^2}\>
\left(\e^{-\nu\alpha}-1\right)   \Theta(k_\perp-\alpha)
\cdot \frac{\as(0)}{4\pi}  , \\
\label{cR12}
 \cR_{12} \!\!&=&\!\!-16C_FC_A\int_0^1 \frac{d\alpha}{\alpha}
\left(\e^{-\nu\alpha}-1\right) 
\!\! 
\int_0^\infty \!\! \frac{d\mu^2\> dk_\perp^2}
 {\mu^2(k_\perp^2+\mu^2)} 
 \ln\frac{k_\perp^2(k_\perp^2\!+\!\mu^2)}{\mu^4}
 \left(\frac{\as}{4\pi}\right)^2 \!\!\Theta(k_\perp\!-\!\alpha). 
 \quad\qquad { }
\emini
The weird factor $\as(0)$ in (\ref{cR11}), which stands for the
two-loop value of the on-mass-shell coupling, 
is there simply because a real gluon has been emitted. 
The contribution $\cR_{12}$ is also ill-defined. 

As we have seen above, $\cR_{11}$,  
is cancelled by the contribution from two-parton gluon splitting 
in the collinear limit in (\ref{beta0}), resulting 
in a finite answer in which $\as(0)$ gets replaced by the running $\as$. 

The second contribution, $\cR_{12}$, taken together with the 
logarithmic piece $\cR_{22}$ of the two-parton emission in $\cR_{2}$,
defined below, produces a finite {\em inclusive correction}\/ to the
answer.

\paragraph{Two-parton emission contribution.}
The second part of the radiator describes gluon decay into two quarks
or into two gluons. 
The latter is described by the non-Abelian part of
the $q\bar{q}\,g\,g$ matrix element $M_{gg}^2$ of (\ref{Mrr2}),
proportional to $C_FC_A$, see [\ref{DMO}].
We have
\beql{cR2}
 \cR_2= 4C_F \int d\Gamma_2\> 
\left(\frac{\as(m^2)}{4\pi}\right)^2 
\left(M_{gg}^2(k_1,k_2)+M_{qq}^2(k_1,k_2)\right) \Omega_T \>.
\eeq
Here essential higher-order virtual corrections have been taken into
account, leading to the running of the coupling with the invariant mass
of the virtual gluon, $|\as(-m^2)|^2=\as^2(m^2)(1+\cO{\as^2})$.  

The ``thrust trigger-function'' $\Omega_T$ in (\ref{cR2})
determines the contribution to the thrust of any  
two-parton configuration and is given by 
\beql{OT}
\Omega_T\>=\> \frac1{2!} \left[\>\exp(-\nu
\min\{\alpha_1,\beta_1\} - \nu\min\{\alpha_2,\beta_2\})
-1 \>\right] .
\eeq
We note that having included the symmetry factor $1/2!$ for both
gluons and quarks we have implicitly defined $M^2_{qq}(k_1,k_2)$ 
to describe the probability of finding {\em either}\/ a quark {\em or}\/ 
an antiquark with momentum $k_1$. 

\paragraph{Splitting the answer.}
Now we are ready to assemble the pieces of the radiator into three 
contributions each of which is explicitly finite.

The ``thrust trigger-function'' $\Omega_T$ in (\ref{cR2}) 
depends on the azimuthal angle between the partons, $\phi$, 
and does not allow the integration in (\ref{beta0}) to be performed.  
It can however be done, 
in a naive (``inclusive'') approximation in which one employs a simplified
version of the $\Omega$-factor, namely,
\beql{Osimp}
  \Omega^{(\mbox{\scriptsize simp})}_T \>=\> \frac1{2!} \left[\>\exp(-\nu
\min\{\alpha,\>\beta\} ) -1 \>\right], 
\eeq 
where $\alpha= \alpha_1 + \alpha_2$ and $\beta = \beta_1 + \beta_2$.
This corresponds to the contribution to the thrust from the parent gluon,  
disregarding the possibility that its offspring may go into opposite
hemispheres. 
The regions $\alpha>\beta$ and $\beta>\alpha$ contribute equally, so
that it is sufficient to consider only the latter:
\beq
  \Omega^{(\mbox{\scriptsize simp})}_T \>\Longrightarrow\>
   \left(\e^{-\nu\alpha}-1 \right)\, \Theta(k_\perp^2+m^2-\alpha^2)\>.
\eeq

Using $\vq{}$ and $\vkp{}$ as integration variables, we obtain
\beql{R2simp}\eqalign{
 \cR_2^{(\mbox{\scriptsize simp})} =& 8C_F 
\int \frac{d\alpha}{\alpha}
\int dk^2_{\perp} \int dm^2
\frac{\alpha^2_s(m^2)}{2!\,(4\pi)^2}
\int_0^1 dz\int_0^{2\pi} \frac{d\phi}{2\pi} \left(M_{gg}^2+M_{qq}^2\right)\>
\Omega^{(\mbox{\scriptsize simp})}_T\>,
}\eeq
with $\phi$ the angle between $\vq{}$ and $\vkp{}$.

The integral in (\ref{R2simp}) has been calculated above in
(\ref{beta0}).
It consists of two terms
$ \cR_2^{(\mbox{\scriptsize simp})}=  \cR_{21} + \cR_{22}$: 
\bminiG{cR2122}
\label{cR21}
 \cR_{21} \!\!\!\!&=&\!\! 8C_F 
\int \frac{d\alpha}{\alpha}  \left(\e^{-\nu\alpha}\!-\!1\right) 
\int \frac{dm^2\>d\kps}{m^2(\kps+m^2)}
\cdot (-\beta_0) \> \left(\frac{\as}{4\pi}\right)^2 \>
\Theta(k_\perp^2+m^2-\alpha^2)  \>; \\
\label{cR22}
\cR_{22} \!\!\!\!&=&\!\! 16 C_FC_A 
\int \frac{d\alpha}{\alpha}  \left(\e^{-\nu\alpha}\!-\!1\right) 
\!\!\int\!\! \frac{dm^2\,d\kps}{m^2(\kps\!+\!m^2)}
\ln\frac{\kps(\kps\!+\!m^2)}{m^4} \left(\frac{\as}{4\pi}\right)^2
\Theta(k_\perp^2\!+\!m^2\!-\!\alpha^2) .
\qquad\quad { }
\emini

Thus we may represent the answer as a sum of three finite terms:
\beql{theanswer}
  \cR= \cR^{(0)} + \cR^{(i)} + \cR^{(n)}\>,
\eeq
where
\bminiG{threeRs}
\label{cR0}
 \cR^{(0)} &=& \cR_{11} + \cR_{21} \>, \\
\label{cRi}
 \cR^{(i)} &=& \cR_{12}+ \cR_{22} \>, \\
\label{cRn}
 \cR^{(n)} &=&  \cR_2 -\cR_2^{(simp)} \>.
\emini
In the first contribution (\ref{cR0}) $\cR_{11}$ involves $\as(0)$
which is cancelled by $\cR_{21}$ to give the standard ``naive''
answer, as we shall see in the next section.
The second contribution, we shall refer to as an ``inclusive
correction''. Here both  $\cR_{12}$ and $\cR_{22}$ contain $m^2\to0$ 
logarithmic singularities which, however, 
are damped by the difference of the two simplified 
``thrust trigger functions'' 
$$
\Omega^{(\mbox{\scriptsize simp})}_T 
- \Omega^{(\mbox{\scriptsize simp})}_T(m^2=0) 
\> \propto \>\Theta(\kps+m^2-\alpha^2)-\Theta(\kps-\alpha^2)\>. 
$$
Similarly, the {\em non-inclusive}\/ correction (\ref{cRn})
is also finite as it involves the difference of the true and the
simplified trigger functions,
\beql{deltaOmega}
\delta\Omega_T = \Omega_T-\Omega^{(\mbox{\scriptsize simp})}_T\>,
\eeq
which also vanishes as $m^2\to0$.

Now we are ready to recall the dispersive method and apply it to 
the calculation of (\ref{threeRs}), starting from the main term
$\cR^{(0)}$. 

\mysection{Na{\"\i}ve power correction for $T$ in the dispersive method}
\subsection{Dispersive method (recollection)}
In the approach of [\ref{BPY}] 
the dispersive representation for $\as$ is implemented,
in which the dispersive variable $m^2$ acts as a ``gluon mass''   
in the matrix elements (see \ref{andim}):
\beql{asae}
 \frac{\as(k^2)}{k^2} =  
 -\int_0^\infty \frac{dm^2}{k^2(k^2+m^2)} \rho_s(m^2) = 
 \int_0^\infty \frac{dm^2}{(m^2+k^2)^2}\ae(m^2)\>.
\eeq
At one-loop level one has
\beql{rhodef}
  \rho_s(m^2) = 
  \frac{d}{d\ln m^2}\>\ae(m^2) \>=\>
  -\frac{\beta_0}{4\pi} \as^2(m^2)\> + \ldots
\eeq
The perturbative answer, $ {\cal{O}}_j(Q^2)$, for an observable $j$
is then formulated in terms of the $m^2$ integral 
\beql{calO}
 {\cal{O}}_j(Q^2) = \int_0^\infty\frac{dm^2}{m^2}
\ae(m^2)\cdot \dot{\cal{F}}_j(m^2/Q^2)\>, 
\quad \dot{\cal{F}}_j \equiv -\frac{d}{d\ln m^2}{\cal{F}}_j\>, 
\eeq
with ${\cal{F}}$ the ``characteristic function'' one derives by
computing the $m^2$-dependent soft-gluon radiation matrix element.
For an infrared/collinear safe  observable 
the $m^2$ integral converges and is mainly
determined by the region $m^2\sim Q^2$,
the typical hard scale of the process. 
This region reproduces the standard
one-loop perturbative prediction 
$$
{\cal{O}}^{\mbox{\small PT}}_j=\ae(Q^2)\,{\cal{F}}(1)
\>\approx\> \as(Q^2)\,{\cal{F}}(1)\>.
$$ 
The genuine non-perturbative component of the answer,
$\delta{\cal{O}}^{\mbox{\small NP}}$, is triggered by the non-perturbative 
component of the effective coupling $\delta\ae(m^2)$.  
Imposing the ITEP/OPE restriction [\ref{OPE}]
that $\as(k^2)$ in the ultraviolet region 
receive no power correction $(k^2)^{-p}$ 
due to the large-distance interaction domain, 
one derives from (\ref{asae}) that the integer moments
of the non-perturbative effective coupling vanish:
$$
 \int_0^\infty \frac{dm^2}{m^2}\delta\ae(m^2)\> (m^2)^i \>=\> 0\>,
 \quad i=1,2, \ldots p\>.
$$
Thus, the power corrections are obtained from the non-analytic 
$m^2\to 0$ behaviour of the one-loop matrix element (characteristic function).
For example, the behaviour 
$\delta{\cal{O}}_T^{\mbox{\small NP}}\sim Q^{-1}$ 
in the thrust distribution originates from
\beql{cFT}
 \dot{\cal{F}}_T(m^2/Q^2) \>\to\> \frac{2C_F}{\pi}
\>\sqrt{\frac{m^2}{Q^2}}\>, \qquad m^2\to 0\>. 
\eeq
The magnitude of the power correction is then expressed in terms 
of the $m$-moments of an ``effective charge''  $\delta\ae(m^2)$
describing the intensity of non-perturbative gluon radiation:
\beq
 \delta{\cal{O}}_j^{\mbox{\small NP}}(Q^2) = 
\int_0^\infty\frac{dm^2}{m^2}
\delta\ae(m^2)\> \dot{\cal{F}}_j(m^2/Q^2) 
\>\simeq\>  C_j \frac{A_{n,q}}{ Q^{n}}\>,
\eeq
where
\beql{momdef}
  A_{2p,q} = \frac{C_F}{2\pi}
  \int_0^\infty \frac{dm^2}{m^2} \>\delta\ae(m^2)\cdot 
  (m^2)^{p}\ln^q{m^2} 
\eeq
(for $p$ half-integer or $q> 0$).
In particular, for the case of thrust from (\ref{cFT}) we deduce: 
$p=\half$, $q=0$, and the
observable-dependent constant $C_j$ is $C_T=4$.

\subsection{Inclusive treatment of the thrust distribution}
In an inclusive approximation one employs the simplified version of the
$\Omega$-factor (\ref{Osimp}).
The contribution $\cR_{21}$  (\ref{cR21}),
proportional to $\beta_0$, is the only one 
which has been dealt with in the past~[\ref{DW97}]. 
Making use of (\ref{rhodef}) 
it can be written as
\beql{cR21mod}
 \cR_{21}\> =\>
 \frac{2C_F}{\pi} 
\int \frac{d\alpha}{\alpha}\left(\e^{-\nu\alpha}-1\right)
\int_0^\infty dm^2 \frac{d\,\ae(m^2)}{d m^2}\>
\int {dk^2_{\perp}} \frac{\Theta(k_\perp^2+m^2-\alpha^2)}
{k_\perp^2+m^2}\>.
 \qquad { }
\eeq
To reconstruct the known result we integrate by parts.
Since $\ae(0)=\as(0)$, the surface term  
cancels $\cR_{11}$, given in (\ref{cR11}),
and we obtain
\beq
\cR^{(0)} \>=\> 
\cR_{11}+\cR_{21} \>=\>  \frac{2C_F}{\pi} 
\int \frac{d\alpha}{\alpha}\left(\e^{-\nu\alpha}-1\right)
\int_0^\infty \frac{dm^2}{m^2} \>\ae(m^2)\>
\Theta(m^2-\alpha^2) \> . 
\eeq
The normal perturbative prediction is obtained by integrating over $\alpha$
between $1/\nu \ll 1$ and $1$. In order to examine the
non-perturbative contribution, one substitutes for $\ae$ its
non-perturbative part, $\delta\ae$, which is concentrated
at small $m^2$. 
This implies that $\alpha$ is small, allowing us to
expand the exponential term,
\beq
\delta\cR^{(0)}=
 -\nu\frac{2C_F}{\pi} 
\int_0^\infty \frac{dm^2}{m^2} \>m\>\delta\ae(m^2)\>
\equiv\> -\nu\> 4A_{1,0}\>,
\eeq
resulting in a non-perturbative shift of the thrust distribution,
proportional to the first moment of $\delta\ae$~[\ref{DW97}],
\beql{shift}
\frac{d\sigma(T)}{dT}
=\frac{d\sigma^{(\mbox{\scriptsize PT})}(T\!-\!\Delta T)}{dT}\>,
\quad \Delta T^{(0)} = \frac{4\,A_{1,0}}{Q}\>.
\eeq
Taking into account subleading $\as^2$-effects, 
the naive inclusive prediction (\ref{shift}) gets modified by the 
{\em thrust rescaling factor}\/  $r_T$:
\beql{shiftmod}
  \Delta T = \frac{4\,A_{1,0}}{Q}\, (1+r_T)\>.
\eeq
This factor originates from the real and virtual two-loop matrix element
and depends on the observable under consideration.

In what follows we analyse and compute the thrust rescaling factor. 
In our framework, it consists of two components:   
an {\em inclusive}\/ $r_T^{(i)}$ and a {\em non-inclusive}\/ 
$r_T^{(n)}$ corrections 

\mysection{Exact treatment}
\subsection{Inclusive correction}
As we have seen above, 
the logarithmic terms in two-particle production $\cR_{22}$
(\ref{cR22}) and
the vertex correction $\cR_{12}$ (\ref{cR12}) cancel in the fully
inclusive anomalous dimension.

In a less inclusive quantity, like thrust, the cancellation is no
longer complete: divergences cancel but a finite correction
remains. 
Indeed, in the real emission contribution the kinematical restriction 
involves the mass of the pair and reads
$$
\Theta\left(\sqrt{\kps+m^2}-\alpha\right)\>.
$$
In the virtual correction, however, we deal with a real gluon restricted
simply by
$$
\Theta\left(\sqrt{\kps}-\alpha\right)\>.
$$
As a result the finite correction which emerges reads
\beql{deltacRi}\eqalign{
 \cR^{(i)} =& 16C_FC_A
 \int\frac{d\alpha}{\alpha}\left(\e^{-\nu\alpha}-1\right)
\int\frac{ dk_\perp^2 \>dm^2}{m^2(k_\perp^2+m^2)} 
 \left(\frac{\as(m^2)}{4\pi}\right)^2  \cr
& \ln\frac{k_\perp^2(k_\perp^2+m^2)}{m^4}
\left[\, \Theta\left(\sqrt{k_\perp^2\!+\!m^2}-\alpha\right) 
- \Theta\left(k_\perp-\alpha \right)\,\right]\,.
}\eeq 
Expanding the exponent and performing the $\alpha$-integration we arrive at
$$\eqalign{
 \cR^{(i)} &= -16C_FC_A\nu \int \frac{dm^2}{m^2} 
 \left(\frac{\as(m^2)}{4\pi}\right)^2  
 \int_0^\infty \frac{d\kps}{\kps+m^2}\left(\sqrt{k_\perp^2\!+\!m^2}
 -k_\perp\right)\> \ln \frac{k_\perp^2(k_\perp^2+m^2)}{m^4} \>,
}$$
where we have extended the $\kps$-integration to infinity since it
converges. 
Now we use (\ref{rhodef})
and integrate by parts to arrive at
\beq\eqalign{
 \cR^{(i)} &= -8C_F\frac{C_A}{\beta_0}\nu \int
 {dm^2}\frac{\ae(m^2)}{4\pi}\> 
 \frac{d}{d{m^2}} \left\{ 2\,m\> c^{(i)}  \right\} 
}\eeq
with
\beq
 c^{(i)}=2 \int_0^\infty \frac{xdx}{1+x^2}\>
 \frac{\ln \,[\,x^2(1+x^2)\,]}{x+\sqrt{1+x^2}}
   = 3.299\>.
\eeq
Substituting $\delta\ae$ for $\ae$ we get the power correction
\beq
  \delta \cR^{(i)} \>=\> -\nu\cdot  c^{(i)} \> 
\frac{2 C_FC_A}{\pi\,\beta_0}\>
 \int \frac{dm^2}{m^2}\>m\> {\delta\ae(m^2)}\>.
\eeq
Recalling the definition of the non-perturbative moment, (\ref{momdef}),
the relative correction to the shift in the thrust distribution
(\ref{shift}) becomes
\beql{DTi}
  \Delta T^{(i)} = \frac{4A_{1,0}}{Q}\cdot r_T^{(i)} \>,\qquad 
r_T^{(i)} = \frac{C_A}{\beta_0}\, c^{(i)} \>=1.100 \>(0.900) 
\quad \mbox{for}\>\> n_f=3\>(0) \>. 
\eeq
Notice that the inclusive correction is of the same magnitude as the 
na{\"\i}ve term. 

\subsection{Non-inclusive correction}
Since the leading power correction originates from the region of 
large gluon radiation angles, $\alpha\sim\beta\sim m$, $k_\perp <m$,
the kinematic region of offspring partons moving into opposite hemispheres
will give an essential correction.

This correction is taken into account by $\cR^{(n)}$ in
(\ref{cRn}), given by the expression (\ref{cR2}) with the
thrust trigger-function $\Omega_T$ replaced by $\delta \Omega_T$
(\ref{deltaOmega}).
In order to obtain the first power correction it suffices to expand 
the exponents in (\ref{OT}) to first order as above.
Straightforward calculation results in
\beql{Otilde}
 \int_0^\infty \frac{d\alpha}{\alpha}\> \delta\Omega
\>\simeq\> -\nu \,\tilde{\Omega}_T\>, \quad 
\tilde{\Omega}_T(q_1,q_2)\>=\> zq_1+(1-z)q_2 - \sqrt{zq_1^2+(1-z)q_2^2}\>.
\eeq
The expression for $\cR^{(n)}$ becomes
\beql{erren}
  \cR^{(n)} \>=\> -\nu\> 4C_F\int dm^2 
\left(\frac{\as(m^2)}{4\pi}\right)^2
\int dk_\perp^2\int_0^1 dz \int\frac{d\phi}{2\pi}
\left(M_{gg}^2+M_{qq}^2\right) \tilde{\Omega}_T\>.
\eeq
Now we use (\ref{rhodef}) and integrate by parts.
Taking into account that at $m\!=\!0$
$$
  \tilde{\Omega}_T(q_1\!=\!q_2) \>=\> 0\>,
$$ 
the surface term vanishes and we get  
\beql{deltacR}
 \cR^{(n)} \>=\> -\nu\> \frac{2C_F}{\pi}\int dm^2 
 \>{\ae(m^2)}\> \frac{d}{dm^2}\left\{
 \int dk_\perp^2 \int dz \int \frac{d\phi}{2\pi} \>
 m^2 \left(M_{gg}^2+M_{qq}^2\right) \frac{\tilde{\Omega}_T}{2\beta_0}
 \right\}
\eeq
The expression in the curly brackets has dimensions of mass.
$Q^2$ enters only in the upper limit
of the $k_\perp^2$ integration which is convergent in the ultra-violet
region.
Therefore we integrate up to infinity so that the result does not depend
on $Q$ and is given by
$$
 \{\mbox{(\ref{deltacR})}\}\>=\>  2\,r_T^{(n)}\,m \>+\> \cO{m^2/Q}\,,
$$ 
with $r_T^{(n)}$ a number still to be calculated.  
In the $m\to0$ limit this leads to
\beq
  \delta\cR^{(n)} \>=\> -\nu\> \frac{4A_{1,0}}{Q}r_T^{(n)}\>.
\eeq
As a result, the non-perturbative shift in the thrust distribution
(\ref{shiftmod}) acquires a contribution 
\beql{shiftR}
  \Delta T^{(n)} \>=\> \frac{4A_{1,0}}{Q}\>r_T^{(n)}\>.
\eeq
In what follows we construct the relevant matrix element 
and compute the non-inclusive component of the thrust rescaling factor
$r_T^{(n)}$.
 
\subsection{Evaluation of $r_T^{(n)}$}
We are now in a position to discuss the determination of the
non-inclusive thrust rescaling factor $r_T$. 
From (\ref{deltacR}), we have that
\beq
 r_T^{(n)} = \frac1{m} \int\frac{d^2k_\perp}{\pi}\frac{d\phi}{2\pi} dz\>
 \frac{\cM^2}{zq_1^2 + (1-z)q_2^2} \>
 \frac{\tilde{\Omega}_T(q_1,q_2)}{4\beta_0}\>.
\eeq
To express the integration measure in terms of 
$u_1$ and $u_2$ we write 
$$\eqalign{
& \frac{d^2k_\perp}{\pi}\frac{d\phi}{2\pi}
 =\frac{d^2k_\perp}{\pi}\frac{d^2q}{\pi}\delta(q^2-(\vq{1}-\vq{2})^2)
 = \frac{d^2q_1}{\pi}\frac{d^2q_2}{\pi}\delta(q^2-(\vq{1}-\vq{2})^2)\cr
 &= {q_1dq_1}\,q_2dq_2\> \frac{2}{\pi}\, d\phi \>\delta(q^2-q_1^2-q_2^2
 +2q_1q_2\cos\phi) 
 = \frac{4q^2}{\pi}\> \frac{u_1du_1\,u_2du_2 }
 {2u_1u_2|\sin\phi|}
 \>. 
}$$
Using the $\delta$-function condition to express $\sin\phi$ we arrive at 
\beq
 \frac{d^2k_\perp}{\pi}\frac{d\phi}{2\pi}
 \>=\> \frac{4q^2}{\pi}\> 
 \frac{u_1du_1 \> u_2du_2}{\sqrt{J}} \>,
\eeq
where the Jacobian factor $J(u_1,u_2)$ is 
$$
 J={((u_1+u_2)^2-1)\,(1-(u_1-u_2)^2)}\>.
$$
Making use of the fact that $\tilde{\Omega}$ in (\ref{Otilde}) satisfies
\beq
 \tilde{\Omega}_T(q_1,q_2) \>=\> q\> \tilde{\Omega}_T(u_1,u_2) 
 \>=\> \frac{m}{\sqrt{z(1-z)}}\> \tilde{\Omega}_T(u_1,u_2)\,,
\eeq
we finally obtain
\beql{rTn}\eqalign{
 r_T^{(n)} =& \frac1{\pi\beta_0}
       \int_0^1 \frac{dz}{\sqrt{z(1-z)}}
       \int_0^\infty u_1du_1 \int_0^\infty u_2du_2 
       \>\frac{1}{\sqrt{J}}\> \cr
 &\times \frac{\cM^2(u_1,u_2)}{zu_1^2 + (1-z)u_2^2} \>\>
  \tilde{\Omega}_T(u_1,u_2)
\>,}
\eeq
where the integrals run over the region $J\ge0$.
This region looks simpler in terms of $u_\pm=u_1\pm u_2$:
\beql{d+-}
\int  \frac{du_1\, du_2}{\sqrt{J}}
\>=\> \half\, \int_{-1}^1\frac{du_-}{\sqrt{1-u_-^2}}
 \int_{1}^\infty\frac{du_+}{\sqrt{u_+^2-1}}\>.
\eeq
\paragraph{Convergence of $r_T^{(n)}$.}
It is necessary to check that the $u_+\to\infty$ and $z(1-z)\to0$ 
regions do not lead to divergences of the $r_T^{(n)}$ integral. 
This is shown in Appendix~B.

Numerical integration results in the value
\beq
   r_T^{(n)}=\frac{2} {\beta_0}\> (-1.227\,C_A +0.365\, C_A \>-0.052\, n_f)\>,
\eeq
where the three terms originate respectively from the soft gluon,
hard gluon and quark matrix elements (\ref{matelms}).
This gives
\beq
  r_T^{(n)}=
     \> -0.710\>(-0.470)\,, \quad \mbox{for}\>\> n_f=3\>(0) \>.
\eeq
We have that the non-inclusive correction alone would lower the ``naive'' 
expectation for the factor in the leading $1/Q$ power correction to
thrust by about 30\%. 
However, 
assembling the inclusive and non-inclusive corrections to the thrust
rescaling factor we finally obtain
\beql{numanswer}
 r_T= r_T^{(i)}+ r_T^{(n)}\>=
\> \frac1{\beta_0} \left[\, 1.575 \,C_A
\>-\> 0.104\,n_f\,\right]\>=\> 0.490\>(0.430)\quad \mbox{for}\>\>
n_f=3\>(0)\,.
\eeq
The conclusion is that the thrust rescaling factor $1+r_T$ 
increases the na{\"\i}ve value by 50\%. The most important correction is the 
inclusive one (\ref{DTi}). 

\mysection{Discussion and conclusions}
Within the dispersive approach of~[\ref{BPY}] power corrections to
perturbative QCD predictions, and to jet-shape observables in
particular, are expressed in terms of moments of the non-perturbative
component of the effective coupling, $\delta\ae(m^2)$.

To fix the absolute normalisation for the leading power corrections to
jet-shape observables a two-loop analysis has to be carried out: since
power corrections are proportional to powers of the QCD scale
$\Lambda$, it is necessary to fix the latter to control the magnitude
of the non-perturbative correction.

In this paper we have examined the thrust distribution in the
high-thrust region to two-loop
accuracy to extract the $1/Q$ power contribution.  To do this it
suffices to consider radiation of soft gluons followed by their
two-parton decays.
 
To this end we have used the physical CMW scheme~[\ref{CMW}] 
in which $\as$ is defined as the strength of inclusive soft-gluon 
radiation (the magnitude of the singular part of the two-loop quark 
anomalous dimension).

Given $\Delta T^{(0)}$ as the result of the na{\"\i}ve inclusive 
one-loop treatment, we have found that the rescaling factor $1+r_T$
for the thrust power correction, $\Delta T=(1+r_T)\Delta T^{(0)}$ is 
rather large. 
This large correction comes mainly from the gluon splitting into two
gluons.

The correction $r_T$ originates from two sources. 
A positive term $r_T^{(i)}$ (the ``inclusive correction'')
comes from an incomplete compensation of the
logarithmic contributions in soft-gluon splitting, 
due to the kinematical difference 
between virtual (massless) and real (massive) gluon radiation
(\ref{deltacRi}). 
This gives the most important correction (see \ref{DTi}).
A negative term $r_T^{(n)}$ (the ``non-inclusive correction'')
is due to the kinematical region of the 4-parton phase space in which the
gluon decay products fly into opposite hemispheres (\ref{erren}).
The contribution to this last term coming from the decay of the 
primary gluon into a quark-antiquark pair proves to be positive and 
numerically very small. 
This agrees with the result of the analysis of Nason and 
Seymour~[\ref{NS}].   
 
It is the region of large gluon emission angle, $\theta\simeq \pi/2$,
that triggers the non-analyticity in the gluon virtuality
$m^2$ and thus power corrections. 
This makes opposite-hemisphere configurations of the decay
products quite common. 
But it so happens that the most important correction is the 
inclusive one. 

Since the moments of $\delta\ae(m^2)$ are not calculable given the
present state of the art, they must be determined phenomenologically.
Therefore to be of any practical value, the programme of calculating
power corrections at two loops must be extended to other observables
driven by the same (first) moment as the thrust.
In a forthcoming publication we will generalise the present analysis
to energy-energy correlations and jet-broadening in 
$e^+e^-$-annihilation~[\ref{futuro}].

For a general observable $j$ we will have a rescaling factor $1+r_j$,
where $r_j$ originates from the primary soft-gluon emission and its
two-parton decay convoluted with the trigger function specific to the
observable $j$. Since $r_j$ depends on the observable, one could argue
that universality of the leading power correction is violated.  On the
other hand, we would prefer to consider universality as still being
valid. Indeed, the ingredients of universality, which are the
universality of soft-gluon radiation and universality of the coupling,
remain intact in the two-loop analysis. Moreover, only at the two-loop
level can universality be given a precise meaning, since only at this
level can one define the QCD scale and thus the coupling. 
The fact that the rescaling factor
is observable dependent does not break universality, since $r_j$
remains under perturbative control.

Let us stress that the two-loop analysis has led to a term $r_T$ in
the rescaling factor which does not contain 
any small parameter and is just a number.
One may wonder whether higher-loop calculations will give corrections
of the same order as $r_j$.
We argue that this is not the case and that terms higher than
two-loops give ``subleading'' contributions, \ie with additional
powers of the perturbative coupling.

To see this observe that the $(\ell+2)$-loop matrix 
element would lead to the appearance of an $m^2$-integral of 
$\as^{\ell+2}(m^2)$.
By using (\ref{rhodef}) and performing the analysis as before, one 
finds that the corresponding contribution to the non-perturbative power 
correction will be proportional to $\ae^{\ell}\delta\ae$ and thus will be 
down, with respect to the two-loop result, by an extra power $\ell$ of 
the coupling.
Strictly speaking, such a perturbative argument is far from being
perfect since the coupling here enters at a small scale. 
Whether the higher order contributions are actually small depends on the
numerical value of the ``perturbative'' effective coupling $\ae/\pi$
at low scales. The large value of the two-loop correction $r_T$
indicates that $\delta\ae$ is {\it smaller}\/ than was indicated by the
naive treatment.

\vspace{5 mm}
\noindent
{\large\bf Acknowledgements}.  We are grateful to A.H.~Mueller and
B.R.~Webber for illuminating discussions.  We also wish to thank
M.~Dasgupta, L.~Magnea and G.~Smye (the authors of [\ref{DasMagSmy}])
and B.R.~Webber for discussions related to the correction of an error
in the first version of this paper, discovered subsequent to an
independent calculation of the $n_f$-dependent piece in
[\ref{DasMagSmy}].
 
\appendix
\mysection{Two-parton production cross section, (\ref{beta0})}
Using the reduced squared matrix element $\cM^2$ in (\ref{cMreduced}),
we can write 
\beql{a1}
 \frac1{2!}\,\int_0^1 dz  \int_0^{2\pi} \frac{d\phi}{2\pi}
\left(M_{gg}^2+M_{qq}^2\right)
= \frac{1}{m^2(k_\perp^2+m^2)} 
\int_0^1 dz \int_0^{2\pi}\frac{d\phi}{2\pi}\;
\left\{C_A\;(2\cS+\cH_g)\;+\;n_f\cH_q\right\}
\,,
\eeq
where the distributions $2\cS$, $\cH_g$ and $\cH_q$ are given in 
(\ref{matelms}) as functions of $\vec{u}_i=\vec{q}_i/q$ with
$$
\vec{q}_1= \vec{k}_{\perp}+(1-z)\vec{q}\,,
\;\;\;\;\;\;
\vec{q}_2= \vec{k}_{\perp}-z\vec{q}
\,.
$$
Performing the integration over $\phi$, the angle between
$\vec{k}_\perp$ and $\vec q$, we obtain
\beql{azim} 
\eqalign{
\int_0^{2\pi} \frac {d \phi}{2\pi}\frac{1}{q_1^2}
&\equiv \;\VEV{\frac{1}{q_1^2}}
\;=\; \frac{1}{\abs{\kps-(1-z)^2q^2}}\,,
\cr
\int_0^{2\pi} \frac {d \phi}{2\pi}\frac{1}{q_1^2q_2^2}
&=\; \frac{1}{m^2+\kps}\VEV{\frac{1-z}{q_1^2}+\frac{z}{q_2^2}}\,,
\cr
\int_0^{2\pi} \frac {d \phi}{2\pi}\frac{q_1^2}{q_2^2}
&=\; 1 \;+\; \VEV{\frac{q^2}{q_2^2}} \;-\; \frac{2}{z}\;B(z)\,,
}
\eeq
where
$$
B(z)\equiv \Theta(z^2q^2-\kps)\;=\;\Theta(z-\frac{\kps}{\kps+m^2})\,.
$$
Similar expressions hold for $\VEV{1/q_2^2}$ and
$\VEV{q_2^2/q_1^2}$.  
By using (\ref{matelms}) we find 
$$
\eqalign{
&
\frac{1}{2!}\VEV{2\cS} \;=\; \frac{1-B(z)-B(1-z)}{z(1-z)}\,,
\cr&
\frac{1}{2!}\VEV{\cH_g} \;=\; -2+z(1-z)
+\frac{B(z)}{2z}+\frac{B(1-z)}{2(1-z)}
+\frac{m^2}{\kps+m^2}\frac{1-6z(1-z)}{2}\,,
\cr&
\frac{1}{2!}\VEV{\cH_q} \;=\; \half - z(1-z)
-\frac{m^2}{\kps+m^2}\frac{1-6z(1-z)}{2}\,.
}
$$
The last two terms in $\VEV{\cH_g}$ and $\VEV{\cH_q}$ vanish upon
$z$-integration and we find 
\beq
\eqalign{
\int_0^1 dz \int_0^{2\pi}\frac{d\phi}{2\pi}\;
&
\left\{C_A\;(2\cS+\cH_g)\;+\;n_f\cH_q\right\}
\;=\;
2 \int_0^1 dz \left\{ C_A(-2+z(1-z))+\frac{n_f}{2}(1-2z(1-z))
\right.
\cr&
\left.
+\frac{C_A}{z}\left(1-B(1-z)-\half B(z)\right)
+\frac{C_A}{1-z}\left(1-B(z)-\half B(1-z)\right)
\right\}
\cr&
=
-\beta_0+ 2 {C_A}\ln\frac{\kps(\kps+m^2)}{m^4}\,.
}
\eeq
This gives (\ref{beta0}).

\mysection{Convergence of $r_T^{(n)}$}
\label{sec:irconv}
To check that the integral (\ref{rTn}) determining $r_T^{(n)}$ converges,
we first consider the large-$u_+$ behaviour of the integrand.
The reduced matrix element (\ref{cMsq}) has a finite limit at
$u_+=\infty$. 
The integration measure in (\ref{d+-}), 
together with the ratio $u_1u_2/(zu_1^2+(1-z)u_2^2)$,
behaves like $du_+/u_+$.
The function $\tilde{\Omega}$ that triggers
the non-inclusive correction to thrust, (\ref{Otilde}), 
can be cast as
\beql{Orev}\eqalign{
 \tilde{\Omega}(u_1,u_2)&=
  \half\left\{ u_+ + (2z-1)u_-  
 - \sqrt{u_+^2 + u_-^2 +2u_+u_-(2z-1)}\right\} \cr
&= \frac{2z(1-z)\,u_-^2}{ u_+ + (2z-1)u_-  
 + \sqrt{(u_+ + (2z-1)u_-)^2 +4z(1-z)u_-^2}} \>.
}\eeq
The crucial point is that it vanishes for large $u_+$,
\beql{Ovan}
 \tilde{\Omega}(u_1,u_2)   
\>\stackrel{u_+\to\infty}{\simeq}\> -\frac{z(1-z)u_-^2}{u_+}\> ,
\eeq
thus ensuring convergence.

By inspecting (\ref{Ovan}) we conclude 
that $\tilde{\Omega}$ also vanishes at $z(1-z)=0$ since
\beql{Oz1z}
 \tilde{\Omega}(u_1,u_2)   
\>\stackrel{z(1\!-\!z)\to0}{\simeq}\> 
\frac{z(1-z)u_-^2}{u_++(2z-1)u_-}\>.
\eeq 
In general, this is sufficient to compensate for the soft singularity of
the matrix elements 
$$
  \cM^2\>\propto\> \frac{1}{z(1-z)}\>.
$$
Special care should be taken when analysing the edge of the phase
space which corresponds to one of the partons being collinear with the
quark direction:
$$
  u_+-1\> \sim\> 1-|u_-| \>\sim\> \sqrt{z(1-z)}\> \ll\> 1 \>, 
$$
where (\ref{Oz1z}) vanishes only as $\sqrt{z(1-z)}$.

For $z \to 0$ the potentially singular region is $u_\pm \to 1$, which
can be explored by writing
$$
u_+=1+(\rho \cos \psi)^2\,,
\;\;\;\;\;\;\;
u_-=1-(\rho \sin \psi)^2\,.
$$
For $z$ and $\rho$ small we have 
$$
zu_1^2+(1-z)u_2^2 = \frac 14 \rho^4 \;+\; z \;+\; \cO{z\rho^2} 
\,.
$$
This combination appears in the denominators of the phase space and
thus generates singularities for $\rho \to 0$ and $z\to 0$.  

Consider first the region $\rho^2  >  \sqrt{z}$.
Here the integrand of $r_T^{(n)}$ has the form
\beql{jk1}dI\; \sim \; \sqrt{z}\;dz\; \frac{d\rho^2}{\rho^4} 
\; \Theta\left(\rho^2-\sqrt {z} \right) \;\; d \psi\;\;  \cM^2
\,.
\eeq
For the soft reduced matrix element $\cS$ one has
$$
\cS \sim \frac{2\rho^2(\cos^2\psi-\sin^2\psi)+\cO{\rho^4} }{z}
\Rightarrow \frac{\cO{\rho^4}}{z}
\,,
$$
where the first term is cancelled upon $\psi$ integration.
The integration over $\rho$ is convergent for $\rho \to 0$ and gives
a contribution with the integrable singularity $1/\sqrt{z}$.

For the hard reduced matrix elements $\cH_g$ and $\cH_q$ 
the dominant contribution in this region is a constant. 
Thus the $\rho$ integration gives a contribution regular in $z$.

Finally, the integration over $\rho$ in the region $\rho^2 < \sqrt{z}$
leads to a contribution which, after angular integration,  is regular
in $z$. 
 
\newpage
\def\labelenumi{[\arabic{enumi}]}

\noindent
{\Large\bf References}
\begin{enumerate}
\item\label{Web94}
        B.R.\ Webber, \pl{339}{148}{94};
       see also {\em Proc.\ Summer School on Hadronic Aspects of
       Collider Physics, Zuoz, Switzerland, 1994} [hep-ph/9411384].
\item\label{StKoZakh}
  G.P.\ Korchemsky and G.\ Sterman, \np{437}{415}{95};\\
    R.\ Akhoury and V.I.\ Zakharov, \pl{357}{646}{95}, \np{465}{295}{96}.
\item\label{DW97}
 Yu.L.\ Dokshitzer and  B.R.\ Webber,  \pl{}{}{97}, to be published.
\item\label{DKT}
    Yu.L.\ Dokshitzer, V.A.\ Khoze and S.I.\ Troyan, \pr{53}{89}{96}.
\item\label{NW}
    P.~Nason and B.R.~Webber, \pl{395}{355}{97}.
\item{\label{BB95}}
   M.\ Beneke and V.M.\ Braun, \np{454}{253}{95}.
\item\label{BPY}
        Yu.L.\ Dokshitzer, G.\ Marchesini and B.R.\ Webber,
       \np{469}{93}{96}.
\item\label{NS} 
       P.\ Nason and M.H.\ Seymour, \np{454}{291}{95}.
\item\label{OPE}
    M.A. Shifman, A.I. Vainstein and V.I. Zakharov, 
    \np{147}{385,448,519}{79}.
\item\label{LBK}
 F.E. Low, \pr{110}{974}{58} ; \\
 T.H. Burnett and N.M. Kroll, \prl{20}{86}{68}.
\item\label{kt}
   V.N. Gribov and L.N. Lipatov, \spj{15}{438, 675}{72}; \\
   D.\ Amati, A.\ Bassetto, M.\ Ciafaloni, G.\ Marchesini
   and G.\ Veneziano,\\ \np{173}{429}{80}; \\
   Yu.L.\ Dokshitzer, D.I.\ Dyakonov and S.I.\ Troyan,
   \prep{58}{270}{80}.
\item\label{lowscales}
    A.C. Mattingly and P.M. Stevenson, \prl{69}{1320}{92}; \\
      G.\ Grunberg, \pl{372}{121}{96}. 
\item\label{DMO}
        Yu.L.\ Dokshitzer, G.\ Marchesini and G.\ Oriani, 
         \np{387}{675}{92}.
\item\label{CWT}
     S.\ Catani, L.\ Trentadue, G.\ Turnock and B.R.\ Webber,
     Phys.\ Lett.\ 263B (1991) 491.
\item\label{CMW}
   S. Catani, G. Marchesini and B.R. Webber,  \np{349}{635}{91}.
\item\label{futuro}
  Yu.L. \ Dokshitzer, A.\ Lucenti,  G.\ Marchesini and G.P.\ Salam,
  under preparation.
\item\label{DasMagSmy} M.~Dasgupta, L.~Magnea and G.~Smye,
JHEP {\bf 9911} (1999) 025.

\end{enumerate}
\end{document}